# A Hybrid Analytical Framework for Asymmetric Pressure and Boundary Layer Wind Simulation in Nor'easters

Seyedeh Fatemeh Mirfakhar[1], Reda Snaiki[1*], Frank Lombardo[2]

[1] *Department of Construction Engineering, École de Technologie Supérieure, Université du Québec, Montréal, Québec, Canada*

[2] *Dept. of Civil and Environmental Engineering, Univ. of Illinois at Urbana-Champaign, Urbana, IL, USA*

[*]*Corresponding author. Email*: reda.snaiki@etsmtl.ca

**Abstract:** Nor'easters frequently impact the North American East Coast, bringing hazardous precipitation, winds, and coastal flooding. Accurate simulation of their pressure and wind fields is essential for forecasting, risk assessment, and infrastructure planning, yet remains challenging due to their complex, asymmetric structure. This study introduces a novel hybrid analytical–data-driven model designed to efficiently simulate Nor'easter pressure and boundary layer wind fields. The pressure field is modeled using an adapted Holland-type formulation, with azimuthally varying parameters estimated through Kriging surrogate models informed by sensitivity analysis of reanalysis data. The wind field is then derived analytically from the momentum equations by decomposing the wind flow into gradient and frictional components. Model performance is assessed against ERA-Interim reanalysis data and surface wind observations from a historical event. Results show that the proposed pressure model accurately reproduces elliptical isobars and key asymmetrical patterns, while the wind model captures the fundamental structure and intensity of the boundary layer flow, including the presence of supergradient winds. Owing to its physical basis, computational efficiency, and ability to represent critical storm asymmetries, the model offers a valuable alternative to computationally expensive numerical simulations for hazard assessment and scenario analysis of extreme Nor'easters.

# 1. Introduction

Nor'easter is a type of intense extratropical cyclone (ETC) that forms along the eastern coast of North America, characterized by powerful winds blowing from the northeast. These coastal storms regularly produce hazardous winds, heavy precipitation, and coastal flooding across the U.S. northeast and Atlantic Canada (Davis & Dolan, 1993; Hall & Booth, 2017; Mäll et al., 2020). Fueled by strong thermal contrasts between cold continental air and comparatively warm maritime air (often enhanced by the Gulf Stream), Nor'easters are most frequent from autumn through spring and can cause prolonged and spatially extensive impacts on infrastructure and communities (Emanuel et al., 1987; Ulbrich et al., 2009; Hall & Booth, 2017; Booth et al., 2018). Accurate and efficient simulation of their pressure and boundary-layer wind fields is therefore critical for forecasting, risk assessment, and engineering design (Befort et al., 2019).

ETCs research has advanced substantially through reanalysis-based diagnostics, objective tracking algorithms, and cyclone compositing techniques that reveal lifecycle and structural patterns (Hodges, 1994; Hoskins & Hodges, 2002; Bauer & Genio, 2006; Catto et al., 2010; Rudeva & Gulev, 2011; Dacre et al., 2012; Neu et al., 2013; Naud et al., 2018; Gramcianinov et al., 2020). Conceptual models such as the Norwegian and Shapiro–Keyser frameworks further clarify typical frontal evolution in these systems (Henry, 1922). Nevertheless, an enduring challenge is representing the pronounced asymmetry characteristic of many Nor'easters: isobars and wind maxima often deviate substantially from the concentric symmetry assumed by many parametric models developed for tropical cyclones (e.g., Holland, 1980). This asymmetry, often tied to frontal structure, storm translation, and environmental shear, critically shapes the spatial distribution of wind hazard (Beare, 2007).

Existing approaches for representing extratropical cyclone winds generally fall broadly into three categories. High-resolution numerical weather prediction (NWP) systems (e.g., WRF) provide detailed physics-based simulations but are computationally expensive and sensitive to resolution and data availability (Nelson et al., 2014; Olson et al., 2019). Empirical and parametric models are computationally efficient but typically assume symmetry or rely on historical regressions that can fail to capture storm-to-storm structural variation (Vickery et al., 1995; Snaiki and Wu, 2018; Done et al., 2020). Analytical boundary-layer models, derived from simplified momentum equations, offer physical insight and computational speed but have rarely been adapted to capture the asymmetric structure of extratropical cyclones (Kepert, 2001; Snaiki & Wu, 2017a, b). Meanwhile, emerging data-driven methods (including knowledge-enhanced machine learning) show promise for parameter estimation but often require large, representative training datasets and careful integration with physical constraints (Snaiki & Wu, 2019, 2022).

Motivated by the need for a computationally efficient yet physically grounded representation of Nor'easter structure, this work develops a hybrid analytical–data-driven model that explicitly represents azimuthal asymmetry in the pressure field and derives the associated boundary-layer wind response. The pressure formulation adapts a Holland-type profile to allow azimuthally varying parameters that describe elliptical isobars and non-uniform radial scales. Key parameters (e.g., azimuthal radius of maximum wind and profile shape) are estimated with Kriging (Gaussian process) surrogate models trained on targeted reanalysis cases, with input predictors selected using global sensitivity analysis. The resulting asymmetric pressure field then drives an analytical boundary-layer solution: the horizontal flow is decomposed into gradient and frictional components by simplifying the momentum equations, yielding a fast, height-resolving wind model that retains core dynamics (Coriolis, pressure gradient, friction) while remaining computationally inexpensive. This work makes three primary contributions: (1) an adapted parametric pressure formulation designed to represent Nor'easter asymmetry; (2) a Kriging-based surrogate framework to predict azimuthally varying pressure parameters from a small set of storm descriptors; and (3) an analytical derivation of the boundary-layer wind field linked to the asymmetric pressure, which is validated using a curated set of Nor'easter cases from the Atlas database (ERA-Interim, 1989–2009) and surface observations from an independent case study to assess near-surface performance.

The remainder of the paper is organized as follows. Section 2 describes the hybrid modeling approach, including the pressure formulation, sensitivity analysis, Kriging implementation, and the analytical wind derivation. Section 3 presents data sources, case selection, and model validation against reanalysis and surface observations. Section 4 discusses results, limitations, and directions for improvement, and Section 5 summarizes conclusions and practical implications.

## 2. Methodology

This study develops a hybrid analytical–data-driven framework to simulate Nor'easter pressure and boundary-layer wind fields. The workflow has three main stages: (a) prescribe a flexible, azimuthally asymmetric parametric pressure field; (b) learn the azimuthal parameter variation from reanalysis using Kriging surrogates whose inputs are selected via global sensitivity analysis; and (c) derive an analytical, height-resolving boundary-layer wind model forced by the reconstructed pressure field. A schematic representation of this overall framework is presented in Fig. 1.

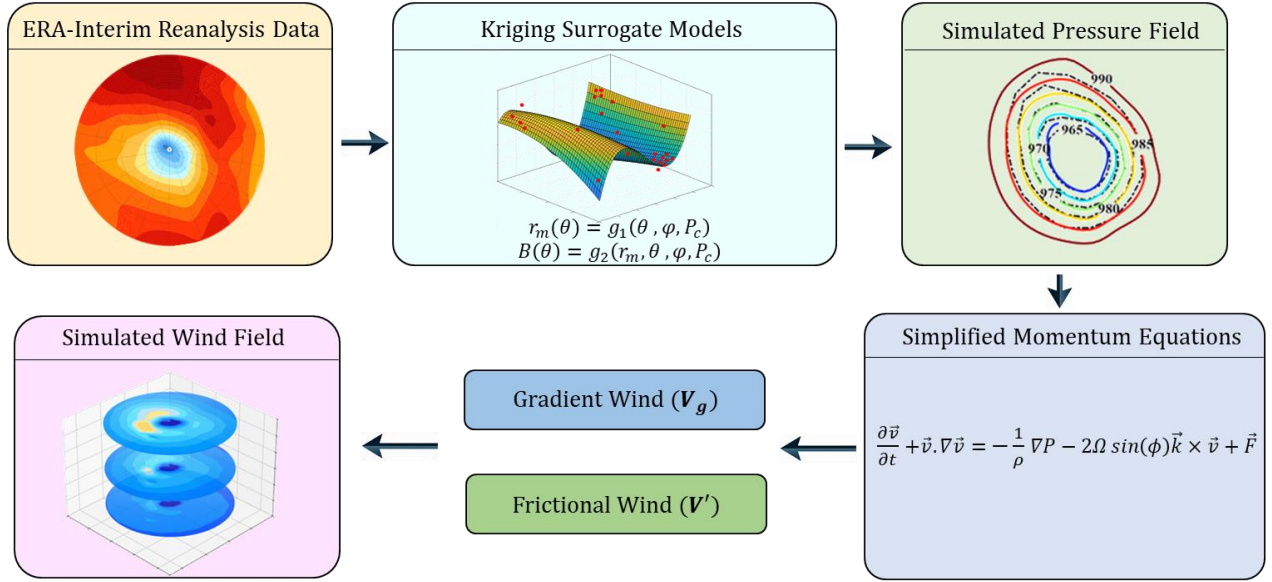

**Fig. 1** Schematic representation of the hybrid analytical–data-driven framework for Nor'easter pressure and wind simulation.

## 2.1 Asymmetric pressure field model

Accurately representing the pressure distribution is fundamental to simulating the wind field. Unlike TCs where isobars are often approximated as concentric circles, Nor'easters exhibit significant asymmetry, although often retaining a generally elliptical or non-circular isobaric pattern. Standard symmetric models, like the original Holland model developed for TCs based on Myers' profile, are therefore insufficient. To capture this asymmetry, this study adopts an adapted pressure field formulation (Snaiki & Wu, 2022):

$$p(r,\theta) = P_c + \Delta p \left\{ \delta\, e^{\left[-\left(\frac{r_m(\theta)}{r}\right)^{B(\theta)}\right]} + (1-\delta)\left(\frac{r}{r_{size}(\theta)}\right)^{n(\theta)} \right\} \quad (1)$$

where $p(r,\theta)$ is the pressure at radial distance $r$ and azimuth $\theta$; $P_c$ is the central pressure; $\Delta p$ is the pressure difference between the storm periphery and the center ($P_n$ - $P_c$, where $P_n$ is ambient pressure); $r_m(\theta)$ is the azimuthally varying radius of maximum winds; $B(\theta)$ is the azimuthally varying pressure profile shape parameter; $r_{size}(\theta)$ is the azimuthally varying distance to the outermost closed isobar (storm size parameter); $n(\theta)$ is another shape parameter (related to the far field); and $\delta$ is a parameter intended to differentiate between storm sectors (e.g., cold vs. warm). It should be noted that detecting distinct sectors ($\delta = 0$ vs $\delta = 1$) in reanalysis data can be challenging. For a simplified parameterization, $\delta = 1$ is assumed for all azimuths. This focuses the model on capturing the primary pressure structure using the exponential term, while asymmetry is primarily encoded within the parameters $r_m(\theta)$ and $B(\theta)$.

## 2.2 Parameter estimation using Kriging surrogate models

The key parameters defining the asymmetric pressure structure in the simplified version of Eq. (1) are the azimuthally varying radius of maximum winds, $r_m(\theta)$, and the shape parameter, $B(\theta)$. To determine these parameters based on the storm's state and specific azimuth, Kriging (Gaussian process regression) is employed to develop surrogate models. This allows the prediction of $r_m(\theta)$ and $B(\theta)$ values at any desired azimuth $\theta$. Two distinct Kriging models are constructed sequentially. The first surrogate model, $g_1$, estimates $r_m$ at a given azimuth $\theta$. The inputs to this model are key storm characteristics – namely, the storm's latitude ($\varphi$) and central pressure ($P_c$) – along with the specific azimuth ($\theta$) itself:

$$r_m(\theta) = g_1(\varphi, P_c, \theta) \tag{2}$$

It should be noted that the surrogate should be applied only within the latitudinal range sampled by the training storms; extrapolation beyond those latitudes may yield unreliable predictions. The second surrogate model, $g_2$, estimates the shape parameter $B$ at the same azimuth $\theta$. This model uses the same set of inputs as $g_1$, and further incorporates the predicted value of $r_m(\theta)$ obtained from the first model:

$$B(\theta) = g_2(\varphi, P_c, \theta, r_m) \tag{3}$$

This sequential Kriging approach allows the shape parameter $B(\theta)$ to depend not only on the overall storm characteristics and azimuth but also on the predicted radius of maximum winds at that specific azimuth. Kriging is advantageous here due to its ability to capture complex, non-linear relationships from data and provide uncertainty estimates for the predictions. The surrogate models ($g_1$ and $g_2$) are trained using a database derived from selected intense, single-cell Nor'easter events in the ERA-Interim reanalysis dataset (1989-2009). The 'target' values for $r_m(\theta)$ and $B(\theta)$ at various azimuths are determined by fitting the pressure model (Eq. 1) or analyzing the storm structure in the reanalysis fields. The corresponding storm characteristics (e.g., $\varphi$ and $P_c$) and the azimuth $\theta$ form the input data for training the Kriging models.

Specific choices regarding the Kriging implementation were made to ensure robust performance. The standard Kriging formulation models the underlying function as a combination of a global trend and a localized stationary Gaussian process capturing deviations from that trend. In this study, polynomial functions were selected to represent the global trend of $r_m(\theta)$ and $B(\theta)$ with respect to their input variables, specifically using a cubic (3$^{rd}$ order) polynomial trend for $r_m(\theta)$ and a quadratic (2$^{nd}$ order) polynomial trend for $B(\theta)$. The spatial correlation structure of the Gaussian process component, which governs the smoothness and local behavior of the surrogate model, was defined using the Matérn 5/2 covariance function. This function is widely used as it offers a good balance between differentiability and flexibility. Crucially, the performance of the Kriging model depends on tuning the hyperparameters associated with the correlation function (e.g., variance and correlation length scales). This tuning was achieved through hyperparameter optimization, performed by maximizing the concentrated log-likelihood function of the training data. A Genetic Algorithm (GA), known for its effectiveness in global optimization searches, was employed to find the optimal hyperparameter values for both the $r_m(\theta)$ and $B(\theta)$ surrogate models.

## 2.3 Analytical wind field derivation

With the asymmetric pressure field $p(r,\theta)$ defined by Eq. (1) and the Kriging-predicted parameters, the wind field $V$ within the boundary layer is derived analytically. The derivation starts from the horizontal momentum equations (simplified from the full Navier-Stokes equations):

$$\frac{\partial V}{\partial t} + V \cdot \nabla V = -\frac{1}{\rho}\nabla P - f\hat{k} \times V + F \tag{4}$$

where $V$ is the horizontal wind vector, $t$ is time, $\rho$ is the air density, $\nabla P$ is the horizontal pressure gradient obtained from Eq. (1), $f = 2\Omega \sin(\varphi)$ is the Coriolis parameter ($\Omega$ is Earth's rotation rate), $\hat{k}$ is the vertical unit vector, and $F$ represents frictional forces. Following common practice for analytical boundary layer models, the wind vector $V$ is decomposed into a gradient wind component $V_g$ (representing the balanced flow above the boundary layer) and a frictional component $V'$ (representing the deviation within the boundary layer):

$$V = V_g + V' \tag{5}$$

### 2.3.1 Gradient wind component ($V_g$)

Above the boundary layer, frictional forces $F$ are negligible. Assuming a steady state in a coordinate system moving with the storm's translation velocity $c$, the horizontal momentum equation for $V_g$ becomes (Meng et al., 1995):

$$(V_g - c) \cdot \nabla V_g = -\frac{1}{\rho}\nabla P - fk \times V_g \tag{6}$$

Solving Eq. (6) in cylindrical coordinates $(r, \theta)$, under the assumption of gradient wind balance (where pressure gradient, Coriolis, and centrifugal forces balance) and neglecting the radial component of the gradient wind ($V_{rg} = 0$) for simplification yields the tangential gradient wind $V_{\theta g}$:

$$V_{\theta g} = \frac{1}{2}[-c\sin(\theta - \theta_0) - 2\Omega \sin(\varphi)\, r] + \left[\left(\frac{-c\sin(\theta-\theta_0) - 2\Omega \sin(\varphi) r}{2}\right)^2 + \frac{r}{\rho} \cdot \frac{\partial P}{\partial r}\right]^{\frac{1}{2}} \quad (7)$$

where $\theta_0$ is the direction of storm motion and $\frac{\partial P}{\partial r}$ is the radial pressure gradient derived from Eq. (1).

### 2.3.2 Frictional wind component ($V'$)

Within the boundary layer, friction is significant. The equation governing the frictional component $V'$ is derived by substituting Eq. (5) into Eq. (4) and subtracting the gradient wind balance Eq. (6). Further assuming a steady state for $V'$ ($\frac{\partial V'}{\partial t} = 0$), neglecting vertical variations in the pressure gradient within the shallow boundary layer, and parameterizing the frictional force $F$ using a constant eddy viscosity $K_m$ (assuming $F = K_m \frac{\partial^2 V'}{\partial z^2}$), leads to simplified equations for the components of $V' = (u, v, w)$ in cylindrical coordinates:

$$u\frac{\partial u}{\partial r} + \frac{v}{r}\frac{\partial u}{\partial \theta} + \omega \frac{\partial u}{\partial z} - \frac{v^2}{r} - fv \approx K_m \frac{\partial^2 u}{\partial z^2} \quad (8)$$

$$u\frac{\partial v}{\partial r} + \frac{v}{r}\frac{\partial v}{\partial \theta} + \omega \frac{\partial v}{\partial z} + \frac{uv}{r} + fu \approx K_m \frac{\partial^2 v}{\partial z^2} \quad (9)$$

where $(u, v, w)$ are the radial, tangential, and vertical components of $V'$, respectively. These equations are solved subject to boundary conditions: 1) $V' \to 0$ as $z \to \infty$ (frictional effect vanishes above the boundary layer); 2) At the surface ($z = z_0$, where $z_0$ is roughness length), a surface friction condition is applied, often using a drag coefficient $C_D$:

$$K_m \frac{\partial u}{\partial z}\bigg|_{z=z_0} = C_D |\mathbf{V}_s| V_r(z_0) \qquad (10)$$

$$K_m \frac{\partial v}{\partial z}\bigg|_{z=z_0} = C_D |\mathbf{V}_s| V_\theta(z_0) \qquad (11)$$

where $\mathbf{V}_s$ is the total wind vector at the surface $z_0$ with $|\mathbf{V}_s| = \sqrt{V_r(z_0)^2 + V_\theta(z_0)^2}$, $V_r(z_0)$ is the total radial wind component at $z_0$ ($V_r(z_0) = V_{rg} + u(z_0)$), and $V_\theta(z_0)$ is the total tangential wind component at $z_0$ ($V_\theta(z_0) = V_{\theta g} + v(z_0)$). Solving the simplified boundary layer equations by linearizing them (i.e., neglecting advection terms in Eq. (8-9)) yields analytical solutions for $u$ and $v$ (the components of $V'$):

$$\begin{bmatrix} u \\ v \end{bmatrix} = e^{-\lambda(z-z_0)} \begin{bmatrix} -\xi & 0 \\ 0 & 1 \end{bmatrix} \begin{bmatrix} D_2 \cos(\lambda(z-z_0)) - D_1 \sin(\lambda(z-z_0)) \\ D_1 \cos(\lambda(z-z_0)) + D_2 \sin(\lambda(z-z_0)) \end{bmatrix} \qquad (12)$$

where $D_1 = -\frac{\chi(\chi+1)V_{\theta g} - \chi V_{rg}/\xi}{1+(\chi+1)^2}$, $D_2 = \frac{\chi V_{\theta g} + \chi(\chi+1)V_{rg}/\xi}{1+(\chi+1)^2}$, $\chi = \frac{C_D}{K_m \lambda}\sqrt{V_{\theta s}^2 + V_{rs}^2}$,

$\lambda = \left(\frac{\partial V_{\theta g}}{\partial r} + \frac{V_{\theta g}}{r} + f\right)^{\frac{1}{4}} \left(2\frac{V_{\theta g}}{r} + f\right)^{\frac{1}{4}} / \sqrt{2K_m}$, and $\xi = 2K_m \lambda^2 / \left(\frac{\partial V_{\theta g}}{\partial r} + \frac{V_{\theta g}}{r} + f\right)$.

## 3. Model Validation and Application

### 3.1 Data and case selection

The model development and validation draw from two complementary datasets: (i) a curated set of intense extratropical cyclones from the Atlas database built from the ERA-Interim reanalysis (Dee et al., 2011; Dacre et al., 2012), and (ii) independent surface observations for a targeted case study used to evaluate near-surface performance. The Atlas database provides objectively tracked

North Atlantic extratropical cyclones identified over 20 winter seasons (December–February) from 1989–2009 (Dacre et al., 2012). ERA-Interim fields at 6-hourly temporal resolution and ≈0.75° horizontal resolution were used as the primary reanalysis source. From the full Atlas catalogue, we selected a subset of 41 storms that exhibit a clear single-cell structure and represent the most intense events in the catalogue. This subset was chosen to (a) focus model training on coherent, well-defined cyclone cores and (b) provide robust azimuthal sampling of pressure structure for the surrogate training. For each selected event we extract the 925-hPa pressure and wind fields, the storm central pressure, and the storm centroid (latitude) from the Atlas metadata. The 925-hPa level was selected because it is the lowest level reliably available in the curated subset and therefore provides the best compromise between representing near-surface forcing and the spatial completeness of the reanalysis fields. For every storm and analysis time step the radial pressure profiles are sampled along a regular set of azimuths (every 10°) and Eq. (1) (Sect. 2.1) is fitted to each azimuthal profile by constrained non-linear least squares to obtain target pairs $\{r_m(\theta_i), B(\theta_i)\}$.

Because the Atlas reanalysis subset contains wind fields only at (and above) 925 hPa, it does not permit a comprehensive direct validation of near-surface winds. To address this gap, we perform a targeted surface-validation case study for the January 4–6, 2018 Nor'easter. Key inputs for this case (storm track and central pressure time series) were obtained from the CANWIN DataHub ETC tracking dataset and associated metadata (Crawford et al., 2020). Observed surface wind gusts at John F. Kennedy International Airport (JFK) and Boston Logan International Airport (KBOS) were retrieved from the Iowa Environmental Mesonet (IEM) station reports. The modeled mean winds at the appropriate reference heights are converted to an equivalent gust proxy (see

Sect. 2.3) to enable direct comparison with station gust observations. Time-series comparisons at these stations evaluate timing, peak magnitude, and event evolution.

**3.2 Pressure field model validation**

This section details the validation of the proposed asymmetric pressure field model. The evaluation encompasses three key aspects: first, the selection and characterization of influential input variables for the parameter estimation models using sensitivity analysis (Section 3.2.1); second, the assessment of the Kriging surrogate models' accuracy in predicting the target parameters $r_m(\theta)$ and $B(\theta)$ (Section 3.2.2); and third, the comparison of the fully reconstructed pressure fields against ERA-Interim reanalysis data (Section 3.2.3).

**3.2.1 Input variable selection via global sensitivity analysis**

To ensure the Kriging surrogate models for $r_m(\theta)$ and $B(\theta)$ (Eqs. 2 & 3) were built using the most relevant predictors, a global sensitivity analysis (GSA) was conducted prior to final model training. This analysis aimed to quantify the influence of potential input storm characteristics on the variability of the target parameters ($r_m$ and $B$, as derived from initial fits to the reanalysis data). The Sobol' method, a variance-based GSA technique, was employed (Saltelli et al., 2008; Snaiki et al., 2024). This method decomposes the output variance into contributions from individual input variables (first-order effects) and their interactions (higher-order effects), with the total Sobol' index representing the overall contribution of each input, including all interactions.

Calculating Sobol' indices typically involves Monte Carlo simulations, which require characterizing the uncertainty or variability of the input parameters. Probability density functions (PDFs) were fitted to the empirical distributions of the potential input variables derived from the 41 selected Nor'easter cases (Section 3.1). The Kolmogorov-Smirnov (KS) test was used to assess the goodness-of-fit for candidate distributions. Figure 2 presents the marginal and bi-dimensional histograms for the characterized variables. For the Monte Carlo simulations required by the Sobol' method, PDFs were fitted to the empirical distributions of the input variables identified in Section 2.2. Based on the KS test, the best-fit distributions used for sampling in the Monte Carlo calculations were: Normal for latitude ($\varphi$), Logistic for central pressure ($P_c$), and Uniform for azimuth ($\theta$). Furthermore, because the prediction of $B(\theta)$ via model $g_2$ uses $r_m(\theta)$ as an input (Eq. 3), the empirical distribution of the target $r_m(\theta)$ values (derived from reanalysis) was also characterized by fitting a Lognormal distribution. This fitted distribution for $r_m(\theta)$ was used when sampling inputs for the sensitivity analysis of $B(\theta)$.

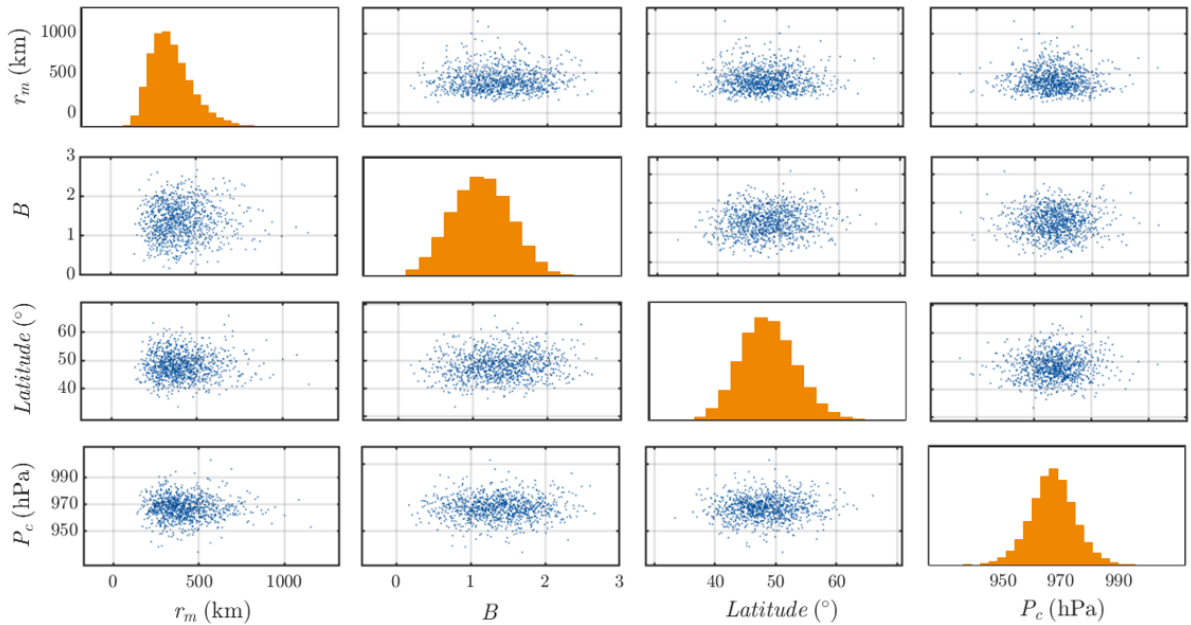

**Fig. 2** Marginal histograms and bi-dimensional scatter plots characterizing key storm variables

Figure 3 presents the total Sobol' indices derived from the sensitivity analysis, quantifying the contribution of each input variable to the variance of the target parameters, $r_m(\theta)$ and $B(\theta)$. The analysis confirmed that latitude, central pressure, and the azimuth all exert a significant influence on the prediction of $r_m(\theta)$. For the shape parameter $B(\theta)$, these same four variables, along with the estimated $r_m(\theta)$ value, were similarly identified as the primary factors driving its variability. Consequently, these influential variables were adopted as the final set of predictors for the Kriging surrogate models ($g_1$ and $g_2$) described in Section 2.2.

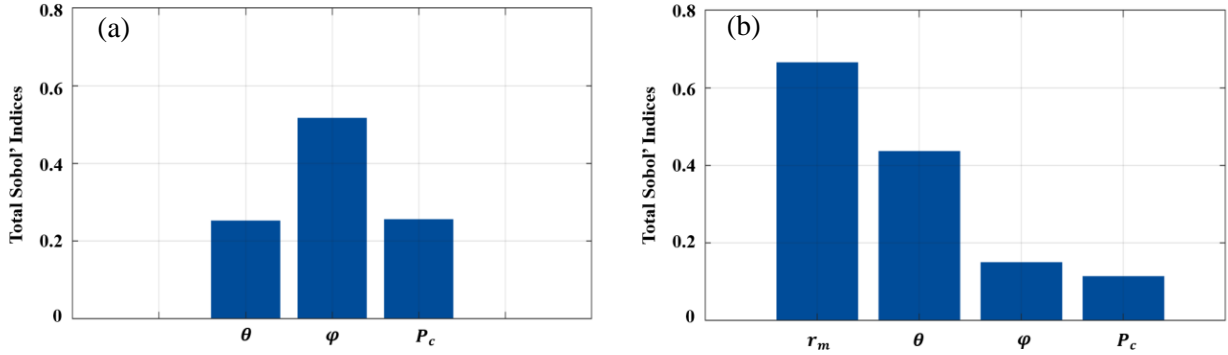

**Fig. 3** Sensitivity analysis results showing total Sobol' indices for the importance of input variables on predicting (a) $r_m(\theta)$ (model $g_1$) and (b) $B(\theta)$ (model $g_2$)

### 3.2.2 Kriging surrogate model performance evaluation

Following the selection of input variables informed by the sensitivity analysis, the performance of the final Kriging surrogate models ($g_1$ for $r_m(\theta)$ and $g_2$ for $B(\theta)$) were rigorously evaluated. The dataset derived from the 41 storm cases was randomly partitioned into an 80% training set and a 20% validation set. Model accuracy was assessed on both sets using the coefficient of determination ($R^2$) and the Root Mean Square Error (RMSE) between the predicted values and the target values obtained from reanalysis data fitting.

For the radius of maximum winds, $r_m(\theta)$, the Kriging model achieved $R^2$ = [0.99] (training) and $R^2$ = [0.98] (validation). The corresponding RMSE was [7.54] km (training) and [18.70] km (validation). Figure 4 provides a visual comparison via a scatter plot of predicted versus target $r_m(\theta)$ values for the training and validation sets, demonstrating a strong clustering around the ideal 1:1 line.

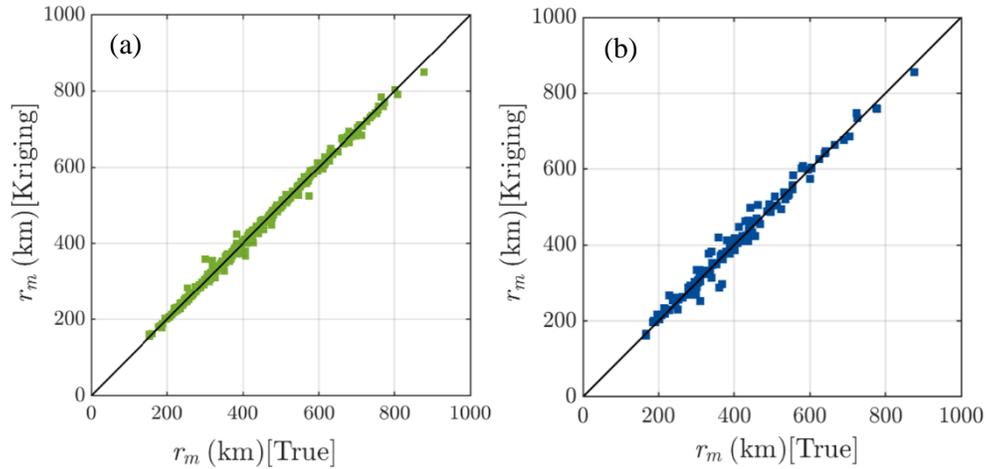

**Fig. 4** Scatter plots comparing Kriging-predicted versus target (reanalysis-derived) radius of maximum wind ($r_m(\theta)$) values for (a) the training set and (b) the validation set

For the shape parameter $B(\theta)$, the model yielded R² = [0.99] (training) and R² = [0.94] (validation). The RMSE values were [0.02] (training) and [0.10] (validation). A similar scatter plot comparing predicted and target $B(\theta)$ values for the training and validation sets are presented in Fig. 5.

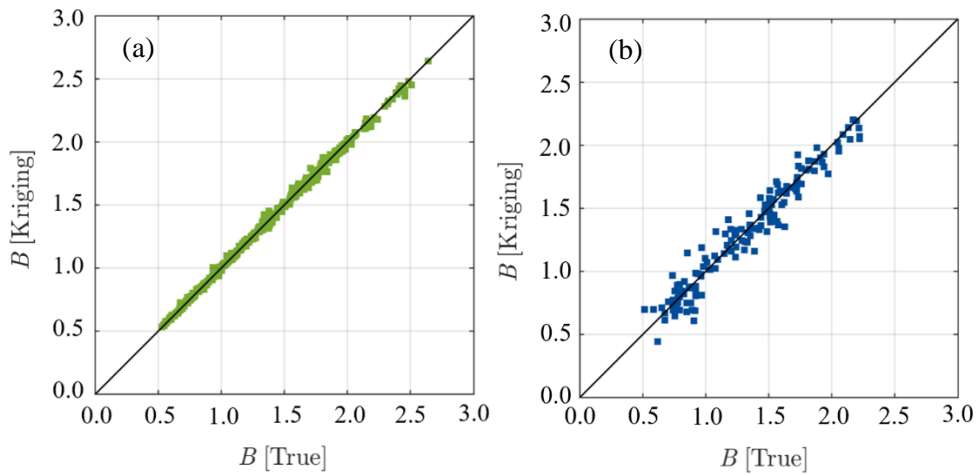

**Fig. 5** Scatter plots comparing Kriging-predicted versus target (reanalysis-derived) shape parameter ($B(\theta)$) values for (a) the training set and (b) the validation set

Furthermore, Fig. 6 compares the PDFs of the target values (derived from reanalysis) against the distributions of the values predicted by the Kriging models, shown separately for the training and validation sets for both $r_m(\theta)$ and $B(\theta)$. The close agreement between the predicted and target distributions in these histograms provides further evidence that the surrogate models effectively capture the statistical characteristics of the parameters.

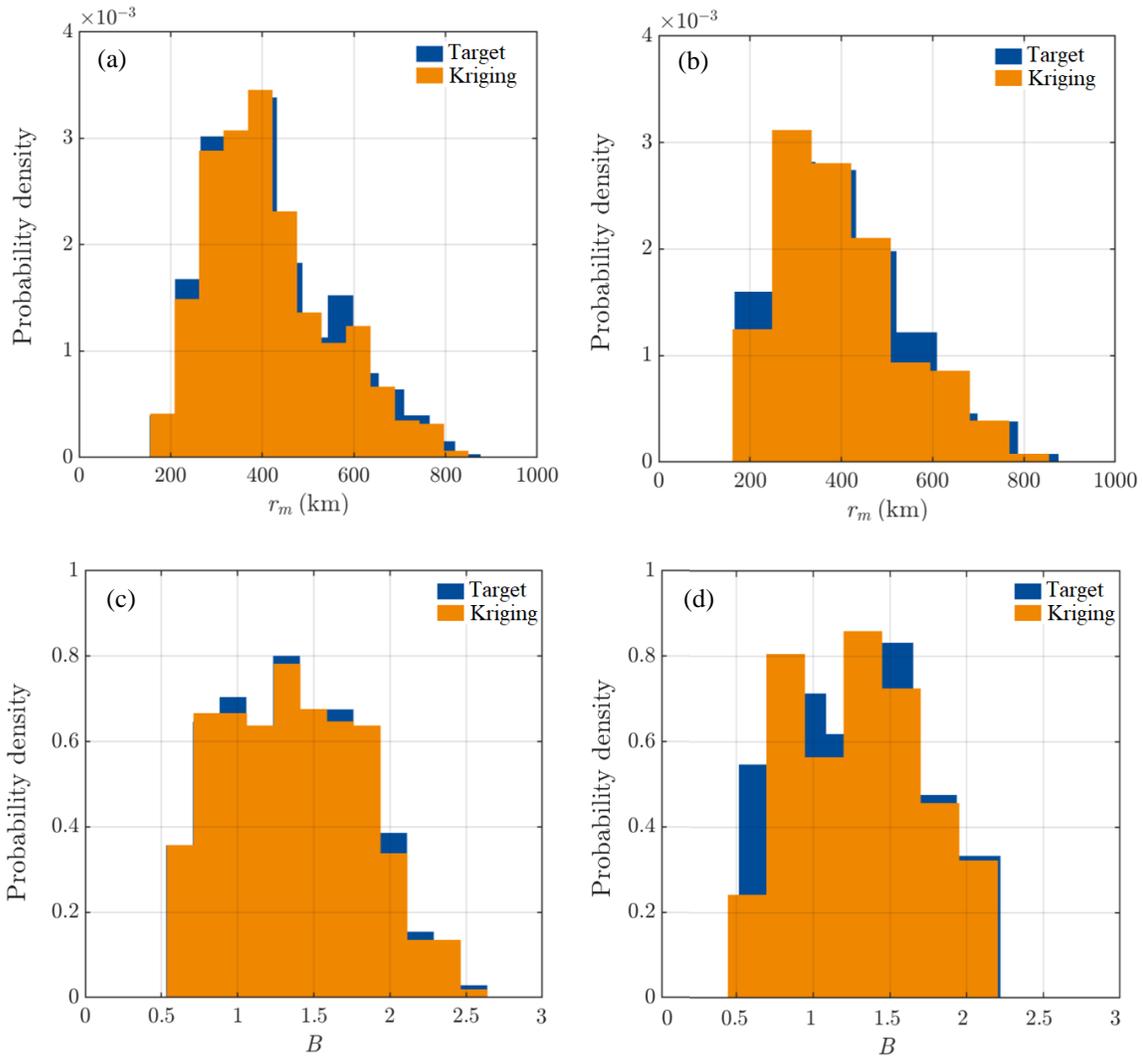

**Fig. 6** Comparison of probability density distributions for target (reanalysis-derived) versus Kriging-predicted values of (a) $r_m(\theta)$ training set, (b) $r_m(\theta)$ validation set, (c) $B(\theta)$ training set, and (d) $B(\theta)$ validation set

Overall, the high correlation coefficients, relatively low RMSE values (considering the inherent variability of these parameters), and good agreement in the probability distributions observed on the independent validation set indicate that the Kriging surrogate models successfully learned the complex underlying relationships. They demonstrate good generalization capability without significant overfitting, providing reliable estimates of $r_m(\theta)$ and $B(\theta)$ for use in the pressure field reconstruction (Eq. 1) and subsequent validation.

**3.2.3 Comparison of modeled pressure fields with reanalysis**

The reconstructed asymmetric pressure fields were evaluated against ERA-Interim 925-hPa analyses for the 41 selected single-cell Nor'easter cases using both qualitative and quantitative diagnostics. Figure 7 presents side-by-side contour comparisons for three representative storms (30 Dec 1989; 12 Feb 2003; 25 Jan 2003). Visually, the modeled fields reproduce the dominant asymmetric features in the reanalysis, including the elliptical isobar patterns and the general orientation of the pressure pattern. Local differences are apparent in storm quadrants where the reanalysis exhibits sharper curvature or more complex sectoral structure (for example, pronounced warm/cold sector contrasts and frontal wrap) that are not fully captured by the simplified formulation used here.

To quantify model skill across the full case set, several metrics were computed comparing the modeled pressure fields ($P_{model}$) with the ERA-Interim reanalysis fields ($P_{reanalysis}$) over the storm domain. The principal summary statistics are a spatial coefficient of determination $R^2$=0.91 and a domain-averaged RMSE = 1.84 hPa. These metrics indicate that the model reliably captures the first-order synoptic patterns and amplitude of the pressure field across the ensemble. The remaining discrepancies can be traced to three main sources. First, the modeling simplification that sets the parameter $\delta = 1$ suppresses explicit warm/cold sector contrasts and therefore limits the ability to reproduce sharp isobar curvature that is often associated with frontal wrapping This behavior is consistent with classical mid-latitude frontal theory (Bjerknes' frontal concept) (Bjerknes and Solberg, 1922), in which distinct warm and cold sectors and associated frontal boundaries produce localized, high-curvature pressure gradients as the cyclone matures and occludes. By enforcing $\delta$ constant the parametric form retains the large-scale azimuthal asymmetry that dominates the pressure signal of many Nor'easters, but it underrepresents small-scale curvature and localized gradient intensification tied to frontal structure or warm-seclusion features. The trade-off was adopted here for parsimony given the limited training sample; however, allowing $\delta$ o vary with azimuth (or using an explicit sector-based decomposition) would enable the model to capture frontal wrapping and sharper isobar curvature; preliminary tests that reintroduce a sectoral multiplier show improved local agreement in some cases (see Sect. 4). Second, uncertainty in the Kriging surrogate estimates of $r_m(\theta)$ and $B(\theta)$ propagates into the reconstructed field; azimuths with sparse or noisy reanalysis profiles exhibit larger surrogate predictive variance and correspondingly larger local pressure error. Third, the spatial resolution and working level of ERA-Interim constrain the representation of fine frontal structure and near-surface gradients and therefore contribute to apparent mismatches in localized gradients. Using higher-resolution reanalyses or additional observational constraints would likely reduce some of these apparent differences.

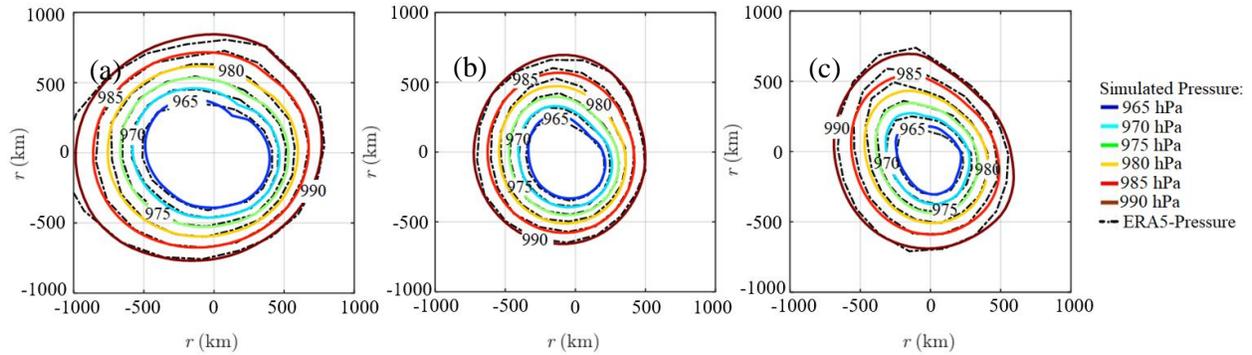

**Fig. 7** Comparison of modeled pressure contours versus ERA-Interim reanalysis pressure contours at the 925 hPa level for three Nor'easter cases: (a) Nor'easter 1 ([30 Dec. 1989]), (b) Nor'easter 2 ([12 Feb. 2003]), and (c) Nor'easter 3 ([25 Jan. 2003])

### 3.3 Wind field model validation

### 3.3.1 Comparison of modeled wind fields with reanalysis

The analytical wind model was evaluated against ERA-Interim 925-hPa wind fields for the three representative Nor'easter cases used in the pressure validation (30 Dec 1989; 12 Feb 2003; 25 Jan 2003), with each storm analyzed at a time when its core was located over the ocean. Because the reanalysis provides winds at 925 hPa but the analytical derivation produces a full height-resolved field, the comparison focuses on the gradient component of the modeled wind at 925 hPa and on bulk spatial metrics that emphasize large-scale pattern and intensity. This choice is consistent with the assumption that frictional effects at this level are reduced and that the gradient wind captures the primary pressure-driven circulation that should be comparable to the reanalysis at that altitude.

Visual comparison of modeled and ERA-Interim wind fields (Fig. 8) shows that the analytical model reproduces the principal synoptic features of Nor'easter circulation. In all three cases the modeled flow exhibits a coherent cyclonic pattern around the low-pressure centre, identifies the approximate radial location of the maximum wind band, and reproduces the general radial decay of wind speed away from the core. The model also provides realistic spatial contrasts between the high-wind region and the surrounding field, reflecting the dominant role of the asymmetric pressure gradient supplied by the Kriging-reconstructed pressure field. Quantitatively, the modeled peak wind magnitudes at 925 hPa compare reasonably with the reanalysis ranges for the three cases considered. For Case 1 (30 Dec 1989) the modeled maximum wind speed was 44.7 m s$^{-1}$, falling within the reanalysis range of approximately 40–45 m s$^{-1}$. For Case 2 (12 Feb 2003) the modeled maximum wind was 44.7 m s$^{-1}$, again near the 40–45 m s$^{-1}$ reanalysis range. For Case 3 (25 Jan 2003) the model produced a maximum of 44.9 m s$^{-1}$, comparable to the reanalysis range of 45–50 m s$^{-1}$. While these per-case maxima indicate the model's ability to approximate synoptic intensity, they do not fully characterize spatial fidelity; differences in the exact shape, extent and local gradients of the high-wind region are apparent.

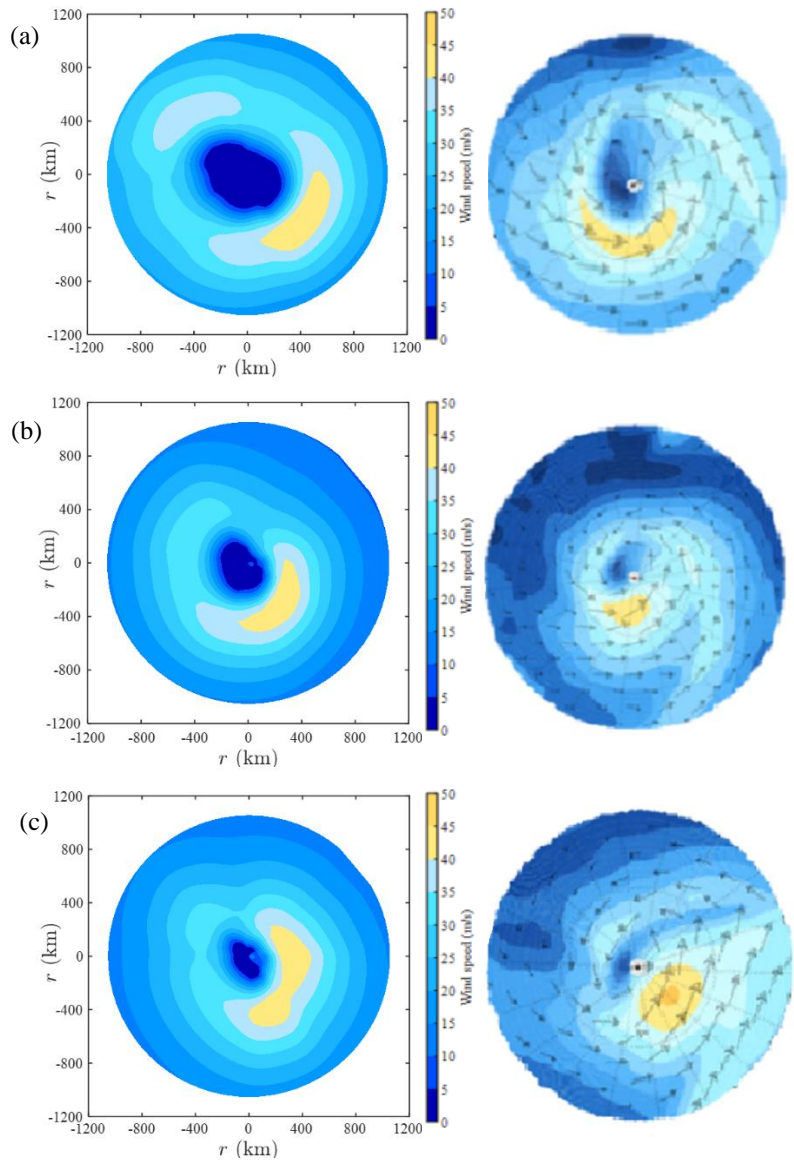

**Fig. 8** Modeled versus ERA-Interim reanalysis wind speed contours (m/s) at the 925 hPa level for Nor'easters 1 (a, [30 Dec. 1989]), 2 (b, [12 Feb. 2003]), and 3 (c, [25 Jan. 2003])

The principal discrepancies between modeled and ERA-Interim winds arise from a combination of model simplifications, surrogate uncertainty, and reanalysis resolution. The analytical formulation omits some transient and small-scale dynamics (for example detailed frontal evolution, transient advection and complex turbulent mixing) that contribute to fine features such as the comma-shaped precipitation/wind patterns present in the reanalysis. In several areas the model shows a rotation or misalignment relative to ERA-Interim (most notably in Case 1), which is attributable in part to the necessary simplifications made in the analytical derivation. Assumptions such as steady-state conditions for the frictional component, the neglect of the radial gradient wind ($V_{rg} \approx 0$), and the use of simplified, constant parameterizations for friction (i.e., $K_m$ and $C_D$) inherently limit the model's ability to capture transient effects and complex turbulent mixing. Furthermore, residual inaccuracies from the input pressure field inevitably propagate into the wind calculations. Finally, the model excludes other important factors known to influence storm asymmetry, such as the environmental vertical wind shear and detailed frontal dynamics.

### 3.3.2 Validation against surface observations

Because the Atlas reanalysis subset used for model development provides wind fields only at and above the 925-hPa level, we evaluated the model's surface-level performance with a targeted case study of the January 4–6, 2018 (Fig. 9) Nor'easter for which high-frequency station observations are available. Storm track coordinates and time-varying central pressure for this event were taken from the CANWIN ETC tracking dataset to drive the pressure reconstruction and the subsequent analytical wind solution. Observed surface wind gusts were obtained from the Iowa Environmental Mesonet (IEM) for two representative coastal airports: John F. Kennedy International Airport (JFK) and Boston Logan International Airport (KBOS). These stations were chosen because they experienced the storm's strongest impacts and provide continuous, quality-controlled gust records suitable for model evaluation.

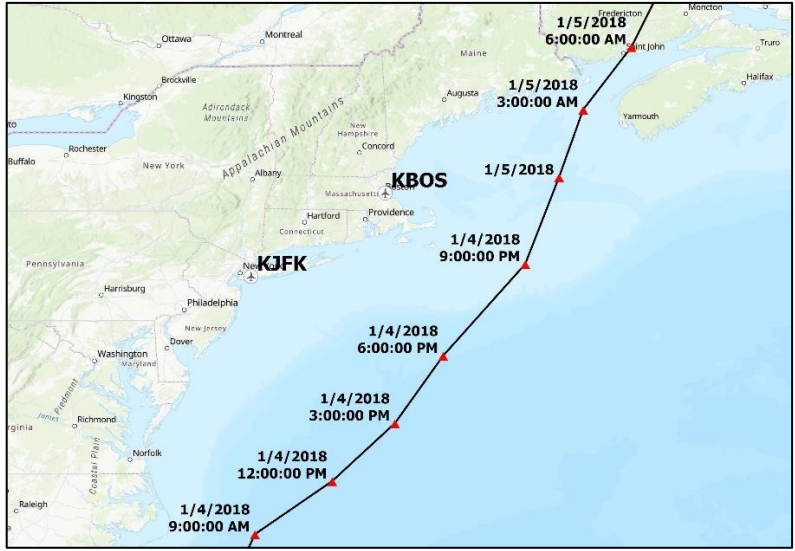

**Fig. *9*** Path of the January 4–6, 2018 Nor'easter used in the validation case study

To facilitate a direct comparison with the observed data, the modeled mean wind speeds were converted to equivalent gust wind speeds. The comparison of the scaled model output and the measured gust speeds at both airport locations is presented in Fig. 10. The model successfully captures the overall trend and timing of the wind event at both sites, including the intensification as the storm approached and the subsequent decrease in wind speeds as it moved away. While some discrepancies in magnitude exist, attributable to the multiple assumptions and simplifications inherent in the model's analytical framework, the results still demonstrate its utility in simulating the wind hazard timeline for a specific, real-world Nor'easter event.

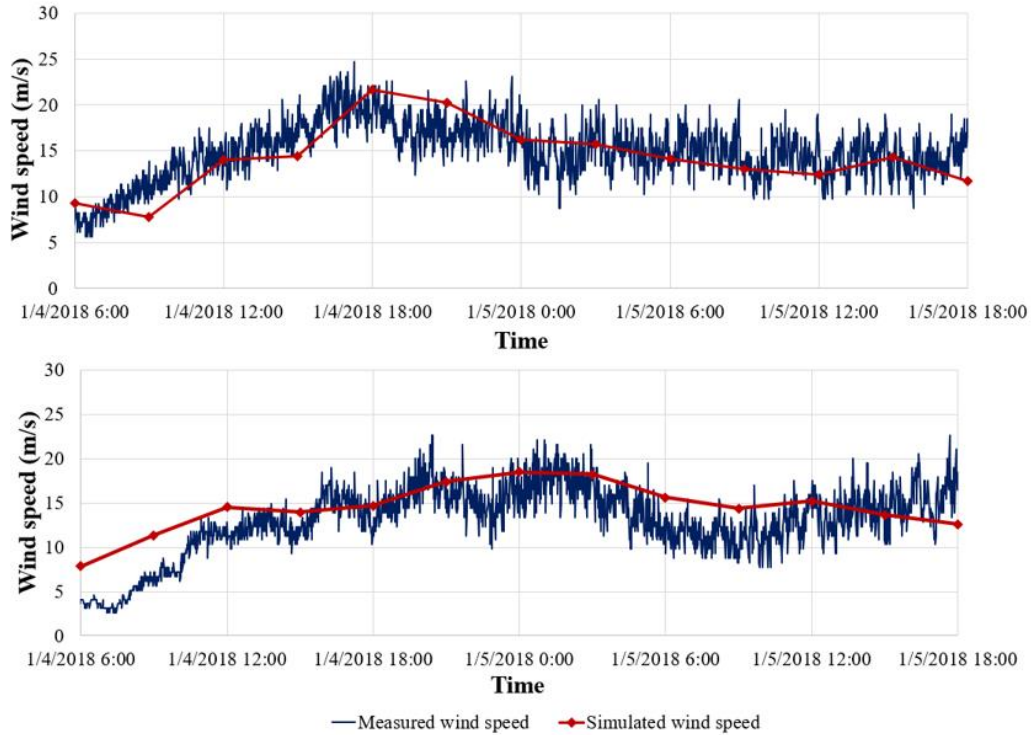

**Fig. 10** Comparison of simulated and measured gust wind speeds for the January 4–6, 2018 Nor'easter at John F. Kennedy Airport (top) and Boston Logan Airport (bottom)

### 3.4 Analysis of Nor'easter boundary layer structure

Building on the demonstrated skill of the model in reproducing the principal features of the asymmetric pressure field and the synoptic-scale wind structure, the analytical wind model is now used to investigate the characteristic structure of the atmospheric boundary layer (ABL) in Nor'easters.

### 3.4.1 Spatial wind field variations with height

To illustrate how the horizontal wind field structure evolves vertically within the ABL, the model was used to generate wind patterns for Nor'easter 1, Nor'easter 2, and Nor'easter 3 at multiple altitudes. Figure 11 displays the modeled wind speed contours at selected elevations, such as near the surface (e.g., 100 m), in the mid-ABL (e.g., 500 m), and near the typical top of the marine ABL (e.g., 1100 m).

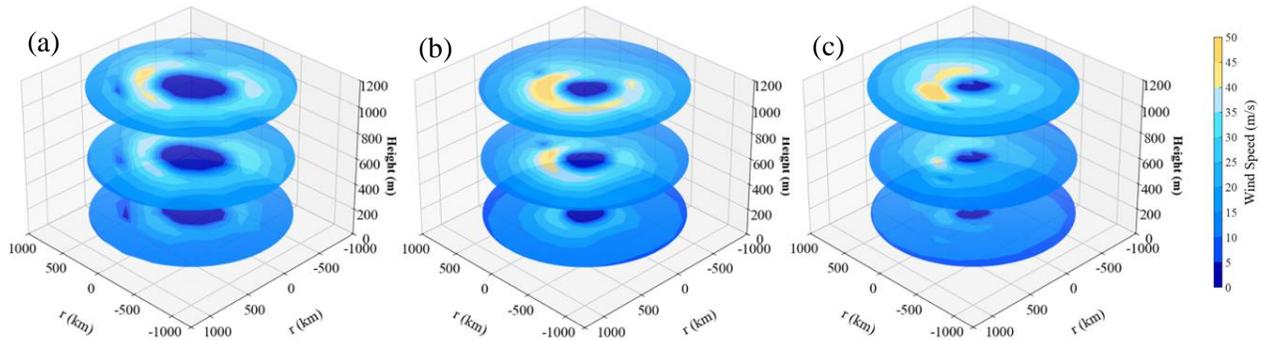

**Fig. 1** Modeled spatial distribution of wind speed (m/s) at 100 m, 500 m, and 1100 m altitudes for three representative Nor'easters: (a) Nor'easter 1 ([30 Dec. 1989]), (b) Nor'easter 2 ([12 Feb. 2003]), and (c) Nor'easter 3 ([25 Jan. 2003])

Analysis of these plots reveals several key features of the modeled boundary layer structure. A distinct increase in the overall wind speed magnitude occurs with increasing height, clearly showing the impact of diminishing surface friction. The spatial patterns also evolve with altitude; for example, the region of maximum wind speed broaden. Also, a characteristic veering of the wind direction with height is observed, indicating stronger inflow angles near the surface compared to the more gradient-parallel flow near the top of the ABL, consistent with Ekman dynamics within a cyclonic system. This analysis underscores the model's ability to represent the three-dimensional interplay between the imposed asymmetric pressure gradient and boundary layer friction.

### 3.4.2 Vertical wind profile analysis at selected locations

For a more detailed view of the vertical structure, wind profiles were extracted from the modeled ABL wind field at several specific locations within Nor'easter 1. These locations, indicated in Fig.

12(a), were chosen to sample different regions relative to the storm center and the maximum wind band (e.g., near core, different azimuths in the max region and periphery). Figure 12(b) presents the vertical profiles of horizontal wind speed from the surface to an altitude above the ABL [e.g., 1500 m] for these points. The modeled profiles exhibit characteristics expected within a turbulent ABL under a strong pressure gradient. Wind speeds typically increase sharply just above the surface due to the high shear induced by surface drag, followed by a more gradual increase with height towards the gradient wind level. The specific shape and magnitude of these profiles are highly dependent on the location within the storm. Profiles located closer to the storm center (e.g., corresponding to the 450 km point) maintain higher wind speeds throughout the ABL depth compared to those at larger radii (e.g., 900 km point), consistent with the radial decay of the pressure gradient force.

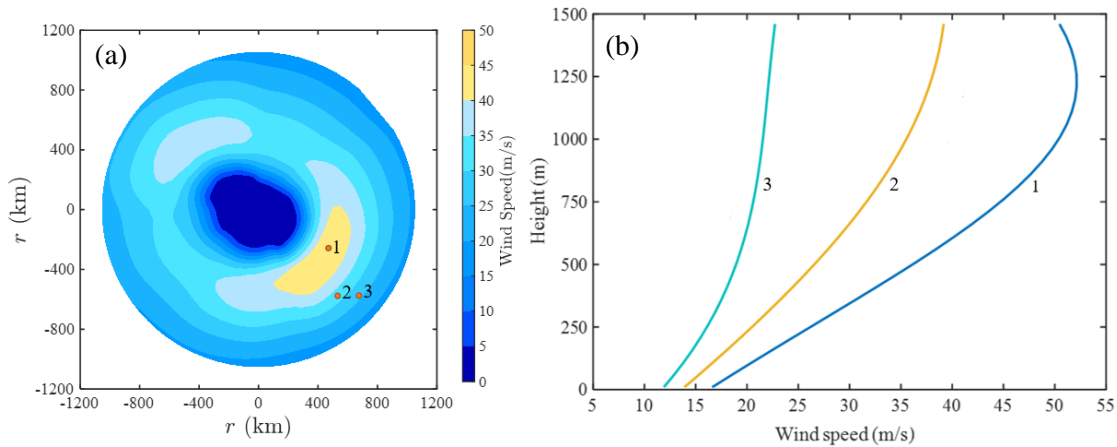

**Fig. 2** Analysis of modeled vertical wind structure for Nor'easter 1 ([30 Dec. 1989]): (a) map showing selected locations (points 1-3) within the modeled wind field at [1500 m], and (b) corresponding modeled vertical profiles of wind speed (m/s) at these locations

A noteworthy feature captured by the model is the presence of super gradient winds within the ABL, particularly in the regions of maximum wind speeds (e.g., corresponding to points 1). As illustrated in Fig. 12b, the wind speed profiles at these locations show a distinct maximum within the ABL (e.g., peaking around 1050-1150 m]), where the speed exceeds the calculated gradient wind speed ($V_{\theta g}$) prevailing above the boundary layer. This phenomenon, often observed in intense cyclones, is physically attributed to the inertial acceleration of air parcels as they are rapidly advected inwards by the strong, frictionally-induced inflow near the surface, conserving angular momentum (Kepert, 2001). The analytical model's ability to simulate this super gradient flow demonstrates its capacity to capture important non-linear dynamics within the boundary layer resulting from the interaction of friction and the pressure gradient force.

## 4. Discussion

### 4.1 Summary of key findings

This study introduced and evaluated a novel hybrid analytical framework for simulating the asymmetric pressure and boundary-layer wind fields of intense Nor'easters. The approach integrates two core components. First, the asymmetric pressure field is prescribed using an adapted parametric formulation whose key azimuthally varying parameters, $r_m(\theta)$ and $B(\theta)$, are efficiently estimated via Kriging surrogate models trained on reanalysis-derived storm datasets. Second, the resulting pressure distribution is used to drive an analytical boundary-layer wind solution, obtained by simplifying the governing momentum equations to retain the essential contributions from pressure-gradient, Coriolis, centrifugal, and frictional forces.

Validation demonstrates the model's skill across multiple domains. When compared with ERA-Interim reanalysis at 925 hPa, the pressure formulation captured the characteristic asymmetry, spatial structure, and amplitude of Nor'easters, achieving high spatial correlations and low RMSE. The analytical wind component reproduced the fundamental cyclonic circulation, the radial location of maximum winds, and the overall magnitude of peak speeds in the reanalysis. In a separate case study of the January 4–6, 2018 Nor'easter, the model captured the temporal evolution and peak timing of observed surface gusts at two coastal stations, with realistic magnitude estimates after applying a site-appropriate gust conversion. Beyond validation, analysis of the model outputs revealed plausible vertical wind profiles, realistic Ekman turning within the lower boundary layer, and the presence of supergradient winds—features consistent with established theory and observational evidence for intense midlatitude cyclones.

The primary significance of this work lies in its ability to represent the asymmetric pressure field—a critical determinant of hazard distribution—in a computationally efficient, physically consistent framework. By leveraging Kriging surrogates, the model can parameterize complex azimuthal structure from a small set of readily available storm characteristics, providing a practical alternative to computationally intensive numerical simulations for applications such as rapid risk assessment and hazard mapping. The physically based wind formulation ensures a direct dynamical link between the pressure and wind fields, preserving essential momentum balances and enabling the emergence of complex features such as supergradient flow without ad hoc tuning. This asymmetric, physics-grounded approach addresses key limitations of purely empirical models, which may lack physical interpretability, and symmetric analytical models, which cannot reproduce the defining spatial patterns of Nor'easters. Overall, the framework provides a robust and efficient tool for simulating the essential pressure and wind-field characteristics of Nor'easters, supporting both scientific understanding and applied hazard assessment.

**4.2 Limitations and future directions**

While the hybrid model demonstrates considerable promise, particularly regarding computational efficiency and the representation of asymmetry, it is essential to acknowledge several limitations inherent in its formulation, the validation data, and the scope of this study. These limitations help contextualize the results and guide future research efforts.

A key simplification impacting the model's fidelity resides in the pressure field formulation. Setting the parameter $\delta = 1$ in Eq. (1) for all azimuths, while simplifying the parameter estimation, effectively neglects the explicit representation of distinct warm and cold thermal sectors characteristic of mature ETCs. This likely contributes significantly to the observed discrepancies between the modeled and reanalysis pressure fields, particularly in the exact isobar shapes and local pressure gradient magnitudes (Section 3.2.2). To investigate this, a preliminary analysis was conducted by applying the generalized form of Eq. (1) (with variable δ) to two additional Nor'easter cases (Case 4: 14 Jan 2002 and Case 5: 21 Dec 1996), using approximated sector information. This information was derived from satellite imagery provided by ERA5-Interim, where the formation of the comma shape indicated the weakening of low-level fronts and the wrapping of the warm front around the cyclone center (Dacre et al., 2012), enabling the estimation of warm and cold fronts. As suggested conceptually in Fig. 13, this generalized approach yielded pressure distributions in closer visual agreement with reanalysis, better capturing structural details. This comparison underscores the impact of the $\delta = 1$ simplification and strongly suggests that future work incorporating objective sector identification or parameterization within the pressure model could lead to significant improvements in accuracy. It is anticipated that such improvements in the pressure field would consequently lead to a more accurate derived wind field.

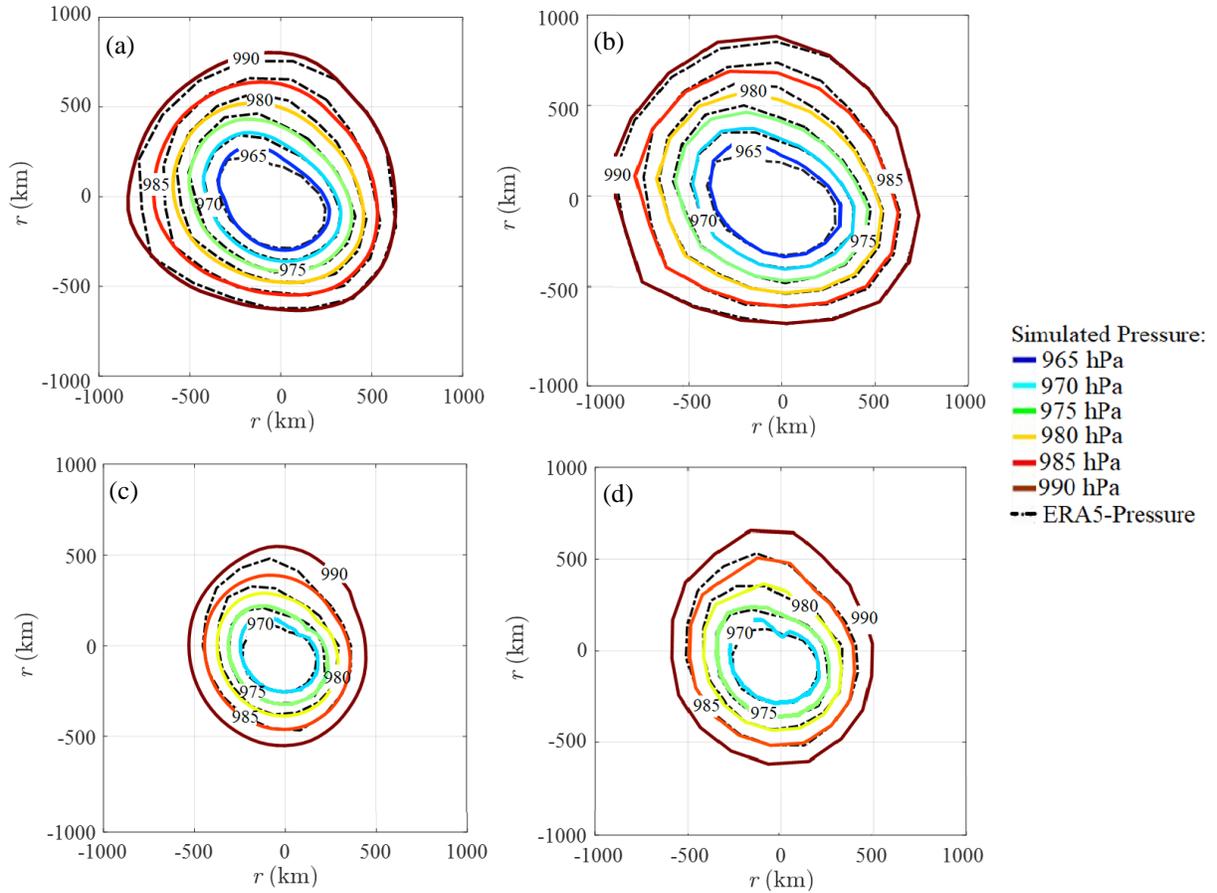

**Fig. 3** Impact of pressure model formulation on simulating 925 hPa pressure fields: comparison with ERA-Interim reanalysis for Nor'easter 4 (a, b) and Nor'easter 5 (c, d) using the simplified ($\delta = 1$) model (a, c) versus the generalized model with estimated fronts/sectors (b, d)

The analytical wind field derivation also involved several necessary simplifications. Assumptions such as steady-state conditions for the frictional component ($V'$), neglecting the radial gradient wind ($V_{rg} \approx 0$), and using simplified constant parameterizations for eddy viscosity ($K_m$) and surface drag ($C_D$) limit the model's ability to capture transient storm evolution, complex turbulent processes, and spatially varying surface interactions. Furthermore, the model excludes important physical processes known to influence Nor'easter structure, including detailed frontal dynamics and interactions with environmental vertical wind shear. Data and validation limitations also warrant consideration. The primary wind field validation against the Atlas database was restricted to the 925 hPa level. While an initial assessment of surface winds was conducted via an independent case study, a comprehensive validation would require testing the model against a much broader set of near-surface observational data. Similarly, the spatial wind field validation at 925 hPa relied on a limited number of case studies. A key question is the generalizability of the proposed framework. The Kriging surrogate models were trained on a curated dataset of intense, single-cell Nor'easter occurring during the winter months. This focus was deliberate, as this category of storm is responsible for the most significant wind and coastal flood damage, whereas less intense events typically have more limited impacts. Consequently, the model's performance is most reliable for this specific class of powerful, well-organized storms. For practical applications, the model is well-suited for storms whose key characteristics, such as central pressure, fall within the range of the training data, allowing for a quick assessment of its suitability. Similarly, the surrogate models should only be applied within the latitudinal range sampled by the training storms, as extrapolation beyond this domain may yield unreliable predictions. A crucial next step, however, is to rigorously establish its applicability across a wider spectrum of conditions. Future work should therefore focus on validating and potentially adapting the model for more diverse events, such as Nor'easters of moderate intensity and storms with complex multi-centered structures. Expanding the framework to other types of extratropical cyclones would further test its general applicability. Moreover, since the proposed model is steady state, it does not simulate the storm's temporal evolution. Many intense Nor'easters are notable for undergoing "explosive cyclogenesis," a form of rapid intensification where the central pressure drops dramatically in a short period. The complex, transient dynamics and strong asymmetries associated with this process, which are inherently captured in reanalysis products, are not represented by the steady-state assumptions of the analytical wind derivation. This likely contributes to some of the observed discrepancies in the wind field's fine structure and intensity when compared to reanalysis, and incorporating parameters related to the storm's intensification rate could be a valuable direction for future research. Finally, the accuracy of the entire framework depends on the Kriging surrogate models, whose performance is tied to the quality of the training data and the chosen input features.

Despite these limitations, the model's efficiency and ability to capture first-order asymmetric features suggest significant potential, particularly for applications in risk assessment where numerous storm scenarios must be simulated rapidly. Future research should prioritize addressing the identified limitations. Implementing and rigorously testing the generalized pressure formula with robust sector identification is a key next step. Refining the analytical wind model by incorporating more sophisticated friction parameterizations or potentially accounting for shear effects warrants investigation. Crucially, validation against more comprehensive datasets, ideally, observational data from buoys, towers, or field campaigns that resolve the boundary layer structure and surface winds, is needed. Performing sensitivity analysis and uncertainty quantification, leveraging the variance information from Kriging, would also strengthen the model's utility. Pursuing these directions could yield a significantly improved and more versatile analytical tool for Nor'easter simulation and impact assessment.

## 5. Conclusion

This study presented a hybrid analytical framework for simulating the asymmetric pressure and boundary-layer wind fields of intense Nor'easters. By combining a physically based parametric pressure model with Kriging-derived azimuthal variability and an analytical boundary-layer wind solution, the approach captures the essential spatial asymmetries and dynamical features that define these storms. Validation against ERA-Interim at 925 hPa demonstrated high spatial correlation and realistic intensity estimates, while a surface-level case study of the January 2018 Nor'easter confirmed the model's ability to reproduce event timing and general magnitude at coastal stations. Analysis of model outputs further revealed plausible vertical wind profiles, realistic Ekman turning, and the occurrence of supergradient winds—features consistent with established boundary-layer theory. The framework's key strength lies in its ability to represent the asymmetric structure of Nor'easters in a computationally efficient yet physically grounded manner. The use of Kriging surrogates enables flexible, data-driven parameter estimation from limited storm characteristics, while the analytical wind derivation ensures a direct dynamical link between the pressure field and resulting winds. This makes the model a practical alternative to full numerical simulations for applications such as rapid hazard assessment, scenario testing, or operational planning. Looking forward, enhancements such as sector-dependent $\delta$ parameterization, higher-resolution surrogate training, and improved boundary-layer formulations could further increase model fidelity. Expansion of validation across a broader set of storms, including other midlatitude cyclone types and tropical–extratropical transition events, would extend its applicability and robustness. Ultimately, the proposed approach offers a versatile and physically interpretable tool for both scientific investigation and applied wind hazard analysis, bridging the gap between simple empirical models and computationally expensive numerical weather prediction systems.

**Acknowledgment:** This work was supported by the Natural Sciences and Engineering Research Council of Canada (NSERC) [grant number CRSNG RGPIN 2022-03492].